\documentclass[12pt]{article}
\usepackage{amsfonts}
\usepackage{amssymb}

\def\hybrid{\topmargin -20pt    \oddsidemargin 0pt
        \headheight 0pt \headsep 0pt
        \textwidth 6.35in       
        \textheight 9.25in       
        \marginparwidth .875in
        \parskip 5pt plus 1pt   \jot = 1.5ex}

\hybrid

\def\baselinestretch{1.2}

\catcode`\@=11

\def\marginnote#1{}
\newcount\hour
\newcount\minute
\newtoks\amorpm
\hour=\time\divide\hour by60
\minute=\time{\multiply\hour by60 \global\advance\minute by-\hour}
\edef\standardtime{{\ifnum\hour<12 \global\amorpm={am}%
        \else\global\amorpm={pm}\advance\hour by-12 \fi
        \ifnum\hour=0 \hour=12 \fi
        \number\hour:\ifnum\minute<10 0\fi\number\minute\the\amorpm}}
\edef\militarytime{\number\hour:\ifnum\minute<10 0\fi\number\minute}

\def\draftlabel#1{{\@bsphack\if@filesw {\let\thepage\relax
   \xdef\@gtempa{\write\@auxout{\string
      \newlabel{#1}{{\@currentlabel}{\thepage}}}}}\@gtempa
   \if@nobreak \ifvmode\nobreak\fi\fi\fi\@esphack}
        \gdef\@eqnlabel{#1}}
\def\@eqnlabel{}
\def\@vacuum{}
\def\draftmarginnote#1{\marginpar{\raggedright\scriptsize\tt#1}}

\def\draft{\oddsidemargin -.5truein
        \def\@oddfoot{\sl preliminary draft \hfil
        \rm\thepage\hfil\sl\today\quad\militarytime}
        \let\@evenfoot\@oddfoot \overfullrule 3pt
        \let\label=\draftlabel
        \let\marginnote=\draftmarginnote
   \def\@eqnnum{(\theequation)\rlap{\kern\marginparsep\tt\@eqnlabel}%
\global\let\@eqnlabel\@vacuum}  }

\def\preprint{\twocolumn\sloppy\flushbottom\parindent 2em
        \leftmargini 2em\leftmarginv .5em\leftmarginvi .5em
        \oddsidemargin -.5in    \evensidemargin -.5in
        \columnsep .4in \footheight 0pt
        \textwidth 10.in        \topmargin  -.4in
        \headheight 12pt \topskip .4in
        \textheight 6.9in \footskip 0pt
        \def\@oddhead{\thepage\hfil\addtocounter{page}{1}\thepage}
        \let\@evenhead\@oddhead \def\@oddfoot{} \def\@evenfoot{} }

\def\numberbysection{\@addtoreset{equation}{section}
        \def\theequation{\thesection.\arabic{equation}}}

\def\underline#1{\relax\ifmmode\@@underline#1\else
        $\@@underline{\hbox{#1}}$\relax\fi}

\def\titlepage{\@restonecolfalse\if@twocolumn\@restonecoltrue\onecolumn
     \else \newpage \fi \thispagestyle{empty}\c@page\z@
        \def\thefootnote{\fnsymbol{footnote}} }

\def\endtitlepage{\if@restonecol\twocolumn \else \newpage \fi
        \def\thefootnote{\arabic{footnote}}
        \setcounter{footnote}{0}}  

\catcode`@=12
\relax

\def\figcap{\section*{Figure Captions\markboth
        {FIGURECAPTIONS}{FIGURECAPTIONS}}\list
        {Figure \arabic{enumi}:\hfill}{\settowidth\labelwidth{Figure
999:}
        \leftmargin\labelwidth
        \advance\leftmargin\labelsep\usecounter{enumi}}}
 \relax
\def\tablecap{\section*{Table Captions\markboth
        {TABLECAPTIONS}{TABLECAPTIONS}}\list
        {Table \arabic{enumi}:\hfill}{\settowidth\labelwidth{Table
999:}
        \leftmargin\labelwidth
        \advance\leftmargin\labelsep\usecounter{enumi}}}
 \relax
\def\reflist{\section*{References\markboth
        {REFLIST}{REFLIST}}\list
        {[\arabic{enumi}]\hfill}{\settowidth\labelwidth{[999]}
        \leftmargin\labelwidth
        \advance\leftmargin\labelsep\usecounter{enumi}}}
 \relax
\makeatletter
\newcounter{pubctr}
\def\publist{\@ifnextchar[{\@publist}{\@@publist}}
\def\@publist[#1]{\list
        {[\arabic{pubctr}]\hfill}{\settowidth\labelwidth{[999]}
        \leftmargin\labelwidth
        \advance\leftmargin\labelsep
        \@nmbrlisttrue\def\@listctr{pubctr}
        \setcounter{pubctr}{#1}\addtocounter{pubctr}{-1}}}
\def\@@publist{\list
        {[\arabic{pubctr}]\hfill}{\settowidth\labelwidth{[999]}
        \leftmargin\labelwidth
        \advance\leftmargin\labelsep
        \@nmbrlisttrue\def\@listctr{pubctr}}}
 \relax
\makeatother
\newskip\humongous \humongous=0pt plus 1000pt minus 1000pt

\newif\ifdtup

\relax

\def\be{\begin{equation}}
\def\ee{\end{equation}}
\def\ba{\begin{eqnarray}}
\def\ea{\end{eqnarray}}

\def\del{\partial}

\def\r{\rho}

\def\G{\Gamma}
\def\d{\delta}

\def\e{\epsilon}

\def\m{\mu}

\def\om{\omega}
\def\Om{\Omega}
\def\l{\lambda}
\def\L{\Lambda}
\def\s{\sigma}
\def\S{\Sigma}

\def\bs{\bigskip}
\def\no{\noindent}

\def\qq{\qquad}

\def\IR{\relax{\rm I\kern-.18em R}}
\renewcommand{\theequation}{\thesection.\arabic{equation}}
\csname @addtoreset\endcsname{equation}{section}

\def \ha {{1\over 2}}

\def \ov {\over}

\def\const{{\rm const.}}

\def\IR{\relax{\rm I\kern-.18em R}}
\def\inv{^{\raise.15ex\hbox{${\scriptscriptstyle -}$}\kern-.05em 1}}

\begin{document}

\newcommand{\beq}{\begin{equation}}
\newcommand{\eeq}[1]{\label{#1}\end{equation}}
\newcommand{\ber}{\begin{eqnarray}}
\newcommand{\eer}[1]{\label{#1}\end{eqnarray}}
\newcommand{\eqn}[1]{(\ref{#1})}
\begin{titlepage}
\begin{center}

\hfill NEIP-04-03\\
\vskip -.1 cm
\hfill hep--th/0406105\\

\vskip .5in

{\Large \bf Supersymmetry and Lorentzian holonomy \\
in various dimensions}

\vskip 0.4in

{\bf Rafael Hern\'andez$^1$},\phantom{x} {\bf Konstadinos
Sfetsos}$^2$\phantom{x}and\phantom{x} {\bf Dimitrios Zoakos}$^2$
\vskip 0.1in

${}^1\!$
Institut de Physique, Universit\'e de Neuch\^atel\\
Breguet 1, CH-2000 Neuch\^atel, Switzerland\\
{\footnotesize{\tt rafael.hernandez@unine.ch}}

\vskip .2in

${}^2\!$
Department of Engineering Sciences, University of Patras\\
26110 Patras, Greece\\
{\footnotesize{\tt sfetsos@des.upatras.gr, dzoakos@upatras.gr}}\\

\end{center}

\vskip .3in

\centerline{\bf Abstract}

\no
We present a systematic method for constructing
manifolds with Lorentzian holonomy group that are non-static
supersymmetric vacua admitting covariantly constant light-like spinors.
It is based on the metric of their
Riemannian counterparts and the realization that, when certain conditions are
satisfied, it is possible to promote constant moduli parameters into arbitrary functions
of the light-cone time. Besides the general formalism, we present
in detail several examples in various dimensions.

\noindent

\vskip .4in
\noindent

\end{titlepage}
\vfill
\eject

\def\baselinestretch{1.2}


\baselineskip 20pt

\section{Introduction}

Ricci-flat manifolds admitting covariantly constant spinors are purely gravitational
supersymmetric solutions of M-theory which imply the existence of
parallel spinors and constrain the holonomy group of the manifold.
In particular, the maximum number of preserved supersymmetries
is given by the number of singlets arising from the decomposition of
the spinor representation
of $Spin(10,1)$ under the holonomy group of the manifold.
When the Killing vector constructed from
the Killing spinor is time-like, the
corresponding spacetimes are static vacua, and
their Riemannian holonomy groups have been completely
classified \cite{Berger}. Non-static
vacua correspond to spacetimes with a covariantly
constant light-like vector, and
although the holonomy groups of Lorentzian
manifolds have not been classified, it is already known
which subgroups of $Spin(10,1)$
leave a null spinor invariant \cite{Bryant}.
Static spacetimes in Euclidean signature satisfying the supergravity
equations of
motion are automatically Ricci-flat.
However, in Lorentzian signature supersymmetric vacua are
only required to be Ricci-null \cite{FOF}, and Ricci-flatness
therefore needs to be
imposed as an additional condition.

Examples of supersymmetric Ricci-null manifolds with Lorentzian
holonomy have been discussed in \cite{Bryant,FOF}, extending a solution already
considered in \cite{Hull}. However, in spite of their great physical interest
there has been no systematic approach to the construction of metrics with Lorentzian
holonomy. In this note we will try to fill this gap constructing  manifolds with reduced Lorentzian
holonomy group in diverse dimensions using the explicitly known form of the metric for some of their
Riemannian counterparts. A possible way to include time-dependence is by allowing the moduli
parameters of a Riemannian metric to be arbitrary functions of the light-cone time.
Although this might seem plausible, it is far from trivial to preserve at
the same time a fraction of maximal supersymmetry and the vacuum
character of the original solution. The present paper is organized as follows:
In section 2 we will develop the general formalism using the supergravity and
Killing spinor equations to obtain $D$-dimensional vacuum supersymmetric
solutions with Lorentzian holonomy group of the semi-direct product type
$\mathbb{G} \ltimes \mathbb{R}^{D-2}$.
In section 3 we will present explicitly
several manifolds with Lorentzian holonomy in
six, eight and nine dimensions, with $\mathbb{G}=SU(2)$, $SU(3)$ and $G_2$,
respectively. To do so it will be enough to utilize a simplified version of our
method. In section 4 we conclude with some remarks and comments on future work.
In an appendix a general six-dimensional solution based on the
four-dimensional self-dual Gibbons--Hawking multi-center metrics
is constructed making use of the method in its full generality.


\section{Ricci-null manifolds}

In this paper we will construct and analyze in detail $D$-dimensional
metrics admitting parallel null spinors of the general form
\be
d\hat{s}_D^2 =  2du\, dv + 2 V_i(u,x) du \, dx^i
+ F(u,x) du^2 + g_{ij}(u,x) dx^i dx^j \ , \:\:\:\:\:
i,j=1,2,\ldots ,D-2\ .
\label{D}
\ee
When the light-cone time $u$ is treated as just a parameter,
the transverse space metrics $g_{ij}(u,x)$
can be taken to be a family of metrics with holonomy
contained in some group $\mathbb{G} \in SO(D-2)$. We will be particularly interested
in cases where $D-2=4,6$ and $7$, with the transverse metrics respectively having an
$SU(2)$, $SU(3)$ and $G_2$ holonomy group.
The functions $V_i$ and $F$ will be determined by requiring that the full metric \eqn{D}
is Ricci-flat and also
preserves a fraction of the sypersymmetry preserved by the transverse metric $g_{ij}$.
Then \eqn{D} will provide non-static M-theory vacua of the form
$ds_{11}^2 = (dx^a)^2 + d\hat{s}_D^2$, where
$a=1,\ldots,11-D$. We will use the frame basis
\be
\hat e^+ = du\ , \qq \hat e^- = dv +V_i dx^i+ \frac{F}{2}du \ ,
\qq \hat e^a= e^a_i(u,x^i) dx^i \ ,
\ee
where $e^a_i$ is the corresponding basis for the transverse metric.
Then,
the only non-vanishing components of the spin connection are
\ba
\hat{\om}^{ab} & \! = \! & \om^{ab}
- \frac{1}{2}\Big(\dot{e}^{[a}e^{b]i}+e^{ai} e^{bj} \del_{[i} V_{j]}\Big)
du\ , \qq  \nonumber \\
\hat{\om}^{a-} & \! = \! & \frac{1}{2}\dot{e}^{a}_{i}dx^i +
\frac{1}{2}\dot{e}^{b}_{i}e^{ai}e_{b} - \frac{1}{2}\left(\partial_{i}F-2 \dot
V_i\right)e^{ai}du  -\ha e^{ai}\del_{[i}V_{j]} dx^j \ ,
\label{kjd0}
\ea
where the dot stands for partial derivatives with respect to $u$,
and $\om^{ab}$ is the spin connection for $g_{ij}(u,x)$. The Killing spinor
equation for purely gravitational backgrounds is
\be
\del_\m \e + {1\ov 4}\hat \om^{AB}_\m \G_{AB}\e = 0 \  , \qq A=(+,-,a)\ .
\label{kjd1}
\ee
We will now assume that the transverse background with metric $g_{ij}(u,x)$,
where $u$ is treated as a parameter, satisfies the Killing spinor equation
\be
\del_i \e + {1\ov 4}\om^{ab}_i \G_{ab}\e = 0 \ ,
\label{kjd2}
\ee
which has non-trivial solutions provided a closed set of projections,
involving products of Gamma-matrices, are imposed on the Killing spinor.
Hence, in this way, a certain fraction of supersymmetry will be
preserved according to the choice of transverse metrics $g_{ij}(u,x)$.
In addition, with the help of these projections the Killing spinor satisfying
\eqn{kjd2} will be left invariant under the action of
the holonomy group of this manifold $\mathbb{G}\in SO(D-2)$.

In order to solve \eqn{kjd1} we further impose the additional projection
\be
\G^+\e = 0 \ ,
\label{uuni}
\ee
and simultaneously require that the deviation of $\hat \om^{ab}$ from
the transverse space connection $\om^{ab}$ is proportional to
a matrix $\L^{ab}$, such that the net result on the Killing spinor is zero.
Namely, we define 
\be
\L^{ab}=\dot{e}_i^{[a}e^{b]i}+e^{ai} e^{bj} \del_{[i} V_{j]}\ 
\label{jdk3}
\ee
and demand that
\be
 \L_{ab}\G^{ab} \e =0\ .
\label{jh9}
\ee
In order for the last condition to be satisfied the left hand side should be
a linear combination of the bilinears in Gamma-matrices that act as
projection operators on the spinor, and guarantee the existence of the Killing 
spinor corresponding to the transverse space metric $g_{ij}$.
The equivalent relation to \eqn{jdk3},
\be
\L_{ab}e^a_i e^b_j = \del_{[i} V_{j]}-\dot{e}_{[i}^{a}e^a_{j]}\ ,
\label{jdk4}
\ee
will also prove useful. Then from the Killing spinor equation
\eqn{kjd1} and the spin connection \eqn{kjd0} we can see that the manifold
(\ref{D}) will preserve half of the supersymmetries preserved by compactifications
along $g_{ij}(u,x)$. The Killing spinor will therefore be
the same as that solving equation \eqn{kjd2}, and in particular $\del_u\e=0$.
We must emphasize that it is highly non-trivial that
we may introduce a $u$-dependence in the transverse metric basis $e^a_i$
such that with an appropriate choice of a set of functions $V_i$ a matrix
$\L^{ab}$ exists, such that the condition \eqn{jh9} is satisfied.

In general, from the integrability condition
of the Killing spinor equation \eqn{kjd1}
it follows that
\be
\hat{R}_{AB} \hat{R}_{AC}\eta^{BC} \e = 0 \ ,\quad \forall \, A\ ,
\label{hjef}
\ee
which, is the condition for a Ricci-null manifold.
Unlike the Euclidean case, in the Lorentzian one this does not automatically
imply Ricci-flatness, i.e. that $\hat R_{AB}=0$, $\forall \, A, B$.
Indeed, specializing in
\eqn{hjef} to the cases $A=a$ and $A=+$ we find that
$\hat{R}_{ab}=\hat{R}_{a+}=0$, but there is no information on
the component $\hat{R}_{++}$.

Returning to our metric ansatz \eqn{D}, the vanishing of $\hat{R}_{ab}$
is apparent from the fact that $\hat \om^{a+}=0$ and that $R_{ab}=0$ by the
assumption \eqn{kjd2} for a supersymmetric transverse metric $g_{ij}$.
However, the vanishing of $\hat R_{a+}$ occurs in a non-trivial way as
it involves properties of the
matrix $\L^{ab}$ and, in particular, \eqn{jh9}.
Hence, we find it worth to present some necessary steps.
First, using the spin connection we obtain the following explicit expression
\be
{\hat R}_{a+} = {\dot \om}^{ab}_i e^i_b +\ha e^i_b\, D_i \L_a{}^b\ ,
\label{dej0}
\ee
where we remind the reader that the covariant derivative 
$D_i \L_a{}^b$ contains the
spin connection $\om^{ab}$.
Then by taking the $u$-derivative of the Killing spinor equation
\eqn{kjd2} we have
\be
{\dot\om}^{ab}_i\G_{ab} \e = 0 \ .
\ee
Let us now contract this relation with $e^{ic}\G_c$, and use $\G_c\G_{ab}=
\G_{cab}-\G_{[a}\d_{b]c}$. This gives
\be
{\dot\om}^{cb}_ie^i_c \G_b \e =-\ha {\dot\om}^{ab}_ie^{ic}\G_{abc} \e =
{1\ov 4} e^{ia} D_i \L^{bc}\G_{abc}\e \ ,
\label{dj9}
\ee
where the last step follows after several algebraic manipulations using the
explicit expression of the spin-connection in terms of the frame
basis components $e^a_i$
together with the condition \eqn{jdk4}. Then we have
\ba
{\hat R}_{a+} \G^a\e & \! = \! & -{1\ov 4} e^{ia}D_i \L^{bc} \G_{abc}\e +\ha e^i_b
D_i \L^{ab}\G_a \e
\nonumber\\
& = \! & -{1\ov 4} e^{ia}\G_a (D_i \L^{bc}\G_{bc}\e)= 0 \ .
\ea
The last equality follows from the Killing spinor equation \eqn{kjd2}
in combination with condition \eqn{jh9}. From the above it easily follows that
${\hat R}_{a+}=0$.

We have seen that Ricci-flatness demands setting the
$\hat{R}_{++}$ component to zero. This will provide a second
order differential equation for $F$, which is found to be
\ba
D_{g}^2F = 2 g^{ij} D_i \dot{V}_j - 2 e^i_a \ddot e_i^a -2 \dot{e}^a_i e^{bi} \Lambda^{ab}
+ \frac {1}{2} \Lambda^{ab} \Lambda_{ab} \ ,
\label{ddFg}
\ea
where we have made use of (\ref{jdk3}), and $D_{g}^2$
is the Laplacian
corresponding to the metric $g_{ij}(u,x)$, in which $u$ is treated as a
parameter.

The holonomy group of the metric \eqn{D} is contained in $\mathbb{G} \ltimes \mathbb{R}^{D-2}$,
which will be generated by bilinears in the Gamma-matrices of the
form $J^{ab}=M^{ab}{}_{cd}\G^{cd}$ and $J^{a}=\G^{a+}$.
The constants $M^{ab}{}_{cd}$ help to
project the generators $\G^{ab}$ of $SO(D-2)$ into the independent generators
$J^{ab}$ of the Lie algebra of $\mathbb{G}$. The $J^{a}$'s
commute among themselves due to the nil-potency of $\G^+$
and form a closed set under the commutator
with $J^{ab}$, with $M^{ab}{}_{cd}$ providing the necessary structure
constants.

The most general way to introduce a time-dependence is to promote the moduli
parameters of the transverse space metric $g_{ij}(u,x)$,
arising as integration constants in solving the Killing spinor
equations \eqn{kjd2}, into arbitrary functions of the variable $u$.
The appearance of the moduli parameters in a solution is ultimately
related to its regularity and to the amount of
symmetry that it preserves. Our method provides a dynamical way
to change the symmetry and singularity structure of a given solution
that is controllable and also consistent with supersymmetry.

There has been in the past works where some solutions of low energy
effective string theory were reinterpreted as if they were originating from solutions
in one less dimension, in which some moduli parameters were made
to depend on the extra dimension (see, for instance,
\cite{Mueller}-\cite{Tseytlin}). Here we have required that the original
solution is supersymmetric and moreover that a fraction of
supersymmetry is preserved when the moduli parameters depend on the extra
coordinates. It is this important extra ingredient that makes it
necessary, as the minimal choice, that the moduli parameters are
functions of the {\it light-cone} time, instead of just the additional ordinary time-like or
space-like coordinate.

When it comes to explicit examples, in this paper we will be
mainly interested in cases where a $u$-dependence
can be introduced in the most minimal way possible. In particular, in cases
where the choice
\be
V_i = 0 \ ,\qq \L^{ab}=0 \ ,
\label{dkj11}
\ee
can be made, and only the function $F$ remains to be determined. Then conditions
\eqn{jdk3}-\eqn{jdk4} simplify to
\be
\dot{e}_i^{[a}e^{b]i} = 0\quad  \Longleftrightarrow \quad
 \dot{e}^a_{[i}e^a_{j]} = 0 \ ,
\label{ee}
\ee
while equation \eqn{ddFg} reduces to
\be
D_{g}^2F = -2 e^i_a \ddot e_i^a \ .
\label{ddF}
\ee
Even within this subclass we may make yet another simplification,
which nevertheless leads to non-trivial results. Namely,
by introducing the $u$-dependence
only in the overall length scale, which then becomes dynamical. In particular,
if $e^a_i(u,x)= \Omega(u)\tilde{e}^a_i(x)$, the condition
\eqn{ee} is trivially satisfied and \eqn{ddF} becomes
\ba
D_{\tilde g}^2F = -2 (D-2) \Om\ddot \Om \ .
\label{ddF1}
\ea
In the appendix we will construct an example where the minimal choice
(\ref{dkj11}) is not valid and instead the method we have presented has to be used
in its full generality.


\section{Supersymmetric waves with Lorentzian holonomy}

For definiteness, we will now present some particular examples
corresponding to manifolds with Lorentzian holonomy group in various dimensions. We will
first construct six-dimensional metrics with $SU(2) \ltimes \mathbb{R}^4$
holonomy, by taking the transverse metric $g_{ij}(u,x)$ to have $SU(2)$ holonomy.
Subsequently, we will construct examples in eight and nine dimensions
having $SU(3) \ltimes \mathbb{R}^6$  and $G_2 \ltimes \mathbb{R}^7$ holonomy,
respectively. Ten-dimensional manifolds with
$SU(4) \ltimes \mathbb{R}^8$ and $Spin(7) \ltimes \mathbb{R}^8$ holonomy
can be constructed
along the same lines, but we will not describe them here.


\subsection{Six dimensions}

Let us first consider a
six-dimensional case preserving $1/4$ of the maximal number of
supercharges. Now $g_{ij}(u,x)$ must be a four-dimensional hyper-K\"ahler manifold,
with $SU(2)$ holonomy group. In the appendix we will consider a general multi-center
solution, but in this section we will simply take the four-dimensional seed solution
with, generically, $SU(2)$ isometry,
\be
ds_{4}^2 = f^2(r) dr^2 + a_1^2(r) \s_1^2
+ a_2^2(r) \s_2^2 + a_3^2(r) \s_3^2 \ ,
\label{SU2s}
\ee
where it is particularly convenient to choose the function $f=a_1a_2a_3$,
and where $\s_i$ are left-invariant Maurer-Cartan $SU(2)$ 1-forms taken to
satisfy
\be
d \s_i = \frac {1}{2} \e_{ijk} \s_j \wedge \s_k \ .
\label{dsigma}
\ee
Using the natural frame
\be
e^i = a_i \s_i\ , \qq i=1,2,3\ , \qq e^4=fdr\
\ee
and imposing on the spinor the projection
\be
\G_4\e = -\G_{123}\e \ ,
\label{pr4}
\ee
one finds that the metric ansatz
\eqn{SU2s} breaks $1/2$ of the maximal supersymmetry and
the functions $a_i$ satisfy a non-linear system of equations
which was first derived by imposing the self-duality condition for the
Riemann curvature for the class of metrics \eqn{SU2s} in \cite{GiPo}.
This is the (Euclidean) Euler system if the Killing spinor is a constant,
and the Halphen system when the Killing spinor depends on the
$SU(2)$ group manifold variables.

Introducing the $u$-dependence we will obtain a six-dimensional
solution with metric
\be
ds_{6}^2 = 2 \, du \, dv + F(u,r) du^2 +
f^2(u,r) dr^2 + a_1^2(u,r) \s_1^2 + a_2^2(u,r) \s_2^2 + a_3^2(u,r)
\s_3^2 \ ,
\label{SU2}
\ee
where we have already restricted in our
ansatz to a function $F$ that is independent of the $SU(2)$ group
manifold variables. Then \eqn{ddF} becomes
\be
F'' = - 4 f^2 \Big( {\ddot{a}_1\ov a_1}+{\ddot{a}_2\ov a_2} +{\ddot{a}_3\ov a_3} +
{\dot{a}_1\dot{a}_2\ov a_1a_2}+ {\dot{a}_1\dot{a}_3\ov a_1a_3} +
{\dot{a}_2\dot{a}_3\ov a_2 a_3}\Big)\ ,
\label{jhd9}
\ee
where the prime denotes the partial derivatives with respect to $r$. In the
simpler case of an overall $u$-dependent length scale, $a_i(u,r)
=\Om(u)\tilde a_i(r)$, the condition \eqn{ddF1} gives
\be
F'' = - 8 \Omega \ddot{\Omega} \tilde{f}^2 \ ,
\label{he1}
\ee
where now $\tilde{f}=\tilde{f}(r)$.

The holonomy group in general is computed by looking at the 
independent generators among
the combinations $\hat R_{AB}{}^{CD}\G_{CD}$ and making sure that they form 
a closed algebra. For the six-dimensional metric \eqn{SU2} 
the holonomy group is found to be
$SU(2) \ltimes \mathbb{R}^4$
and the explicit form of the independent generators is
\be
J^{i} = \G^{4i}-\ha \e_{ijk}\G^{jk}\ ,\qq i=1,2,3\ ,\qq \G^{a+}\ ,\qq
a=(i,4) \ .
\ee
Note that \eqn{pr4}, together with the universal projection \eqn{uuni},
restricts the Killing spinor to be an invariant of the holonomy group.

Among the most important examples that can be constructed in this case are those involving
the Eguchi--Hanson and the Taub--NUT metrics.

\subsubsection{Eguchi--Hanson}

In this case we have \cite{EH}
\be
a_1^2(r)=a_2^2(r)= m^2 \hbox
{coth} (m^2r) \ , \qq a_3^2(r)=\frac{2m^2}{\sinh(2m^2r)} \ , \qq r\geq 0\ ,
\ee
where $m$ is the moduli parameter.
At $r\to \infty$ there is a removable singularity of the {\it bolt}
type, and the
asymptotic region is located at $r\to 0$.
The lift to eleven
dimensions of the static limit of this manifold is
$ds_{11}^2=dx_{1,6}^2 + ds_{EH}^2$, and corresponds to a solution
describing a set of flat D6-branes.
When the constant moduli parameter $m$ is replaced by a function
$m(u)$ we have from \eqn{jhd9} that
\ba
F(u,r) & = & 2 \dot m^2 \left[ \coth(m^2 r) + {m^2r\ov \sinh^2(m^2r)}
-2 m^4r^2 {\coth(m^2r)\ov \sinh^2(m^2r)} \right]\
\nonumber\\
&&+\ 2 m \ddot m \left[ {m^2r\ov \sinh^2(m^2r)}-\coth(m^2r)\right]\ ,
\qq m=m(u)\ ,
\ea
where a function of $u$ coming from the integration has been set to zero in order to avoid a
linearly increasing behaviour of $F(u,r)$ as $r \to \infty$.
The six-dimensional metric is regular since for large $r$, near the
{\it bolt}, the function $F\simeq 2\dot m^2 -2 m \ddot m$ and therefore
it can be absorbed by a shift of the coordinate $v$. Hence, the six-dimensional
space near $r\to \infty$ looks like $S^2 \times \mathbb{R}^2$ times the
two-dimensional light-cone. In the asymptotic region at $r\to 0$ the function
$F\simeq -{4\ov 3} m^3\ddot m r \to 0$, and we get the four-dimensional
Euclidean space, as a cone over $S^3/{\mathbb Z}_2$, times the two-dimensional light-cone.
Hence we interpolate between these two spaces of different topology
as the variable $r$ exhausts
its full range from $r=0$ to $r\to \infty$. This interpolation can also happen
in a dynamical way if we choose for the function $m(u)$ the following
specific behaviour in the remote past and future for the light-cone time $u$,
\be
\lim_{u\to -\infty}m =0 \ ,\qq \lim_{u\to \infty}m \to \infty\ .
\ee
At the end points the symmetry group for the metric \eqn{SU2} gets enhanced, as
a result of its evolution in time (light-cone).
Another possibility is to have functions behaving as
$\lim_{u\to \pm\infty}m =m_\pm$. In that case we obtain in the remote past and
remote future the direct product of the four-dimensional Eguchi--Hanson metric with
the two-dimensional light-cone.\footnote{Similar remarks can be made for the other explicit
examples we present in this paper.}

For the simpler situation where the $u$-dependence is via an overall
function in the metric,
$\Om=\Om(u)$, we easily conclude from \eqn{he1} that
\be
F(u,r)= -4 \Om \ddot \Om \, m^2 \coth(m^2 r)\ , \qq m=\const\ ,
\ee
where again a function arising from the integration has been chosen such that there is
no growing
behaviour in the limit $r\to \infty$. As before near the {\em bolt} the
solution looks like $S^2 \times \mathbb{R}^2$ times the
two-dimensional light-cone. In the asymptotic region $r\to 0$ the function
$F$ blows up as $1/r$.

\subsubsection{Taub--NUT}

The Taub--NUT metric corresponds to \cite{Taub-NUT}\footnote{
The holonomy of the purely geometrical solution 
constructed by combining a wave with the Taub--NUT space 
\cite{intersections1,intersections2} has been 
worked out in \cite{Batrachenko}.}
\be
a_1^2(r)=a_2^2(r)= {1+4 m^2 r\ov 4 m^2 r^2} \ ,
\qq a_3^2(r) = \frac{4 m^2}{1+4 m^2 r} \ , \qq r\geq 0\ ,
\ee
where $m$ is the moduli parameter. The removable
{\it nut} singularity is located at $r\to \infty$, and the asymptotic
region at $r\to 0$. For the case where the constant moduli parameter $m$
is replaced by a function $m(u)$ we conclude from \eqn{jhd9} that
\ba
F(u,r) = {m\ddot m-3 \dot m^2\ov 6 m^4 r^2} \ , \qq m=m(u)\ .
\ea
When the $u$-dependence is via an overall function in the metric, from
\eqn{he1} we obtain
\be
F(u,r)= -\Om \ddot \Om \left({4\ov r}
+{1\ov 3 m^2 r^2}\right) \ ,\qq m= \hbox{constant} \ ,
\ee
where again in both cases a function of $u$ arising from the integration 
has been chosen such that there is no growing behaviour as $r\to \infty$.
In both cases the function $F$ vanishes near the {\it nut} singularity as
$r\to \infty$, and blows up asymptotically when $r\to 0$.


\subsection{Eight dimensions}

A family of metrics with $SU(3) \ltimes \mathbb{R}^6$ holonomy,
preserving $1/8$ supercharges, arises when $g_{ij}(u,x)$ is a
Calabi-Yau threefold with $SU(3)$ holonomy group. The
six-dimensional seed solution that we will employ has the form
\be
ds_6^2 =
f^2(r) dr^2 + a_1^2(r)(\s_1^2 + \s_2^2) + a_2^2(r) (\hat{\s}_1^2 +
\hat{\s}_2^2) + b^2(r) (\s_3 + \hat{\s}_3)^2 \ ,
\label{hfew}
\ee
where $\s_i$ and $\hat{\s}_i$, with $i=1,2$, are now triplets of Maurer--Cartan
1-forms on a two-sphere, normalized so that they again obey
(\ref{dsigma}). Using the natural frame
\be
e^{i}=a_{1}\s_{i}\ ,\quad {e}^{\hat i}=a_{2}\hat{\s}_{i}\ , \quad i=1,2\ , \qq
e^{3}=b(\s_3 + \hat{\s}_3)\ ,\qq e^{4}=fdr\ ,
\label{hej3}
\ee
and imposing the two projections (see, for instance, \cite{EN})
\be
\G_{12}\e =\hat \G_{12}\e = -\G_{34}\e \ ,
\label{pr8}
\ee
one finds a set of differential equations for the coefficients, and
the metric \eqn{hfew} breaks $1/4$ of the maximal supersymmetry.
The explicit solution is the so-called resolved conifold metric
\cite{conifold} with
\ba
a_1^2(r) \! & = \! & \frac {r^2+6a^2}{6} \ ,
\qq a_2^2(r) = \frac {r^2}{6} \ , \nonumber \\
f^2(r) \! & = \! & \frac {r^2 + 6a^2}{r^2+9a^2} \ , \qq
b^2(r) = \frac {r^2}{9 f^{2}(r)} \ .
\label{conifold}
\ea
The limit $a \rightarrow 0$ corresponds to the singular conifold.
The $u$-dependent eight-dimensional metric will
then be
\ba
ds_8^2 & \! = \! & 2 \, du \, dv + F(u,r) du^2 + f^2(u,r) dr^2 + a_1^2(u,r) ( \s_1^2 + \s_2^2 )
\nonumber \\
&&\phantom{} + a_2^2(u,r) (\hat{\s}_1^2 + \hat{\s}_2^2) + b^2(u,r)
(\s_3+\hat{\s}_3)^2 \ ,
\label{SU3}
\ea
where again the function
$F$ is taken to be independent of the $SU(2)$ group manifold variables.
From the condition (\ref{ddF}) we now get the differential equation
\be
\left(f^{-1}{a_1^2 a_2^2 b }F^\prime\right)^\prime
=- 2 f a_1^2 a_2^2 b\Big( \frac {\ddot{f}}{f}
+ 2 \frac {\ddot{a}_1}{a_1} + 2 \frac {\ddot{a}_2}{a_2}
+ \frac {\ddot{b}}{b} \Big)  \ .
\label{ddFSU3}
\ee
When the resolved conifold
moduli is replaced by a $u$-dependent function, $a=a(u)$,
we obtain from (\ref{ddFSU3}) and (\ref{conifold})
the explicit solution
\be
F(u,r) = - 27 \frac {a^2 \dot{a}^2}{r^2+9 a^2}- 3 (\dot a^2 + a\ddot a)
\ln(r^2+9a^2) \ ,
\ee
where a function of $u$ arising from the integration has been chosen so that 
there is no $1/r^2$ term. 
  
When the $u$-dependence comes through a conformal factor,
the differential equation for $F$ \eqn{ddF1} simplifies to
\be
\left(f^{-1}{a_1^2 a_2^2 b }F^\prime\right)^\prime
=- 12 \Omega \ddot{\Omega} f a_1^2 a_2^2 b \ .
\ee
Using (\ref{conifold}) we can solve for $F$ to obtain
\be
F(u,r) = - \Om \ddot{\Om} r^2 - \frac {c(u)}{18a^2r^2} + \frac {c(u)}{162a^4}
\ln \Big( 1 + \frac {9a^2}{r^2} \Big) \ ,
\ee
where $c(u)$ is a function arising from the integration.

The holonomy group of the metric \eqn{SU3} is $SU(3) \ltimes
\mathbb{R}^6$, and the explicit form of the independent generators is
found to be
\ba
J_1 & = & \G_{\hat{1}2}+\G_{1\hat{2}} , \qq
J_2=\G_{2\hat{2}}+\G_{1\hat{1}} , \qq
J_3=\G_{\hat{1}\hat{2}}-\G_{12} , \qq  \nonumber \\ J_4 & = &
\G_{42}+\G_{13} , \qq J_5=\G_{23}-\G_{41} , \qq
J_6=\G_{3\hat{1}}-\G_{4\hat{2}} ,\qq \\ J_7 & = &
\G_{4\hat{1}}-\G_{\hat{2}3} , \qq
J_8=3^{-1/2}(2\G_{43}-\G_{12}-\G_{\hat{1}\hat{2}}) \ ,\nonumber
\ea
together with $\G^{a+}$. We can easily verify that the projections
\eqn{pr8} and \eqn{uuni} guarantee that the Killing spinor is an invariant of the holonomy group.


\subsection{Nine dimensions}

Let us now take $g_{ij}(u,x)$ to be a seven-manifold of $G_2$ holonomy,
with an $SU(2) \times SU(2)$ isometry. The seed solution will be the seven-dimensional
metric ansatz
\be
ds_{7}^2 = dr^2 + \sum_{i=1}^3 a_i^2(r) \s_i^2
+ \sum_{i=1}^3 b_i^2(r) \Big( \S_i + c_i(r) \s_i \Big)^2 \ ,
\label{G23}
\ee
where the $\s_i$'s and $\S_i$'s are two different sets of $SU(2)$
Maurer--Cartan 1-forms obeying the normalization condition \eqn{dsigma}.
Using the natural frame
\be
e^{i}=a_{i}\s_{i}\ ,\quad {e}^{\hat i}=b_i (\S_i + c_i\s_i)
\ ,  \quad e^7= dr\ , \quad i=1,2,3\ , \quad \hat i = i +3\ ,
\label{hej9}
\ee
and imposing the three independent projections (see, for instance,
\cite{EN}-\cite{EPR1})
\be
\G_{ij}\e =-\G_{\hat i\hat j}\e \ ,\qq \G_7\e=
-\G_{\hat 1 \hat 2\hat 3}\e \ ,
\label{pr9}
\ee
one finds a first order system of coupled differential equations for the coefficients, and
the metric \eqn{hfew} breaks $1/8$ of the maximal supersymmetry.
The functions $c_i$ are not
independent: they are given as rational functions of the $a_i$'s and $b_i$'s
(for details see equations (30) and (31) of \cite{HS}).
In particular, a useful property of the $c_i$'s is that they do not scale
when all functions $a_i$ and $b_i$ are scaled by the same factor.
Allowing now all functions to depend on the light-cone time $u$ we obtain an M-theory vacuum
preserving $1/16$ supercharges, with non-trivial nine-dimensional metric
with $G_2 \ltimes \mathbb{R}^7$ holonomy group given by
\be
ds_{9}^2 = 2 \, du \, dv + F(u,r) du^2 + dr^2 + \sum_{i=1}^3 a_i^2(u,r) \s_i^2
+ \sum_{i=1}^3 b_i^2(u,r) \Big( \S_i + c_i(u,r) \s_i \Big)^2 \ .
\label{G2}
\ee
However, applying condition \eqn{ee} to this metric we get the constraint $\sum_{i=1}^3 {a_i\dot c_i
\ov b_i}=0$, which is satisfied when $\dot c_i=0$, for $i=1,2,3$.
In particular we can make the functions $c_i$ independent of $u$ by allowing
only $u$-dependence as an overall factor in the metric, because, as mentioned 
before, they do not scale when we scale all the coefficients $a_i$ and
$b_i$ by the same factor. An exception to this is the round case where all $a_i$ are equal
to each other, and similarly for the functions $b_i$, since then $c_i=-1/2$, for $i=1,2,3$.
This is the manifold we will consider in detail.\footnote{It would be
interesting to
investigate if the general metric \eqn{G23} is consistent with supersymmetry
and remains a vacuum solution by employing the general construction via
\eqn{D} and \eqn{jdk3}, that has non-vanishing $V_i$ and $\L^{ab}$.
On the contrary, the metrics
in \cite{BGGG} with an $SU(2) \times SU(2) \times \mathbb{Z}_2$
isometry, although not in the class of metrics \eqn{G23},
can be employed to construct,
within the minimal choice \eqn{dkj11}, a nine-dimensional manifold with
Lorentzian holonomy.}

In the round case the isometry group of $g_{ij}$
develops an additional $SU(2)$ factor, and the form of
the coefficients in (\ref{G2}) is explicitely known \cite{BS}. In this case,
promoting the moduli parameter into a function of $u$, we have
\be
ds_{9}^2 = 2 \, du \, dv + F du^2 +
\Big( 1 - \frac {\r_0^3(u)}{\r^3} \Big)^{-1} d\r^2
+ \frac{\r^2}{12} \sum_{i=1}^3 \s_i^2 + \frac{\r^2}{9} \Big( 1 - \frac {\r_0^3(u)}{\r^3}
\Big)\sum_{i=1}^3 \Big( \S_i - \frac{1}{2} \s_i \Big)^2 \ ,
\ee
and from equation (\ref{ddF}) we can solve for $F$. We find that
\be
F' = \frac {6\r^4}{(\r^3-\r_0^3(u))^2} \,
\del_u \Big( \r_0^2 \, \dot{\r}_0 \Big) + {c(u)\r^3\ov (\r^3-\r_0^3(u))^2}\ ,
\label{FG2}
\ee
where $c(u)$ arises from the integration. The form of $F(u,\r)$ can be readily obtained from
\eqn{FG2}, but it is a complicated expression and we will not present it here.

If we consider instead
the case where all $u$-dependence in the metric is through a conformal factor
we find that
\be
F(u,r)=-{7\ov 12} \Om\ddot \Om \left[3\r \left(\r+{2\r_0^3\ov \r^2
+\r_0\r+\r_0^2}\right) -4 \sqrt{3} \r_0^2 \tan\inv \left(2\r+\r_0\ov
\sqrt{3}\r_0\right)\right]\ ,
\ee
where an integration constant has been chosen such that $F$ remains finite as
$\r\to \r_0$.

We finally note that the generators of the holonomy group
$G_2 \ltimes \mathbb{R}^7$, labeled according to \eqn{hej9}, are
given by
\ba
&& \G_{ij}+\G_{\hat i\hat j}\ ,\qq \G_{i\hat j}-\G_{\hat i j}\ ,
\nonumber\\
&& 2 \G_{7i}+\e_{ijk}\G_{j\hat k}\ ,
\qq  2 \G_{7\hat i}+\ha \e_{ijk}(\G_{j k}-\G_{\hat j \hat k})\ ,
\label{g22g}
\\
&& \G_{1\hat 1}-\G_{2\hat 2}\ ,\qq \G_{2\hat 2}-\G_{3\hat 3}\ .
\nonumber
\ea
together with $\G^{a+}$. The generators of
the $G_2$ algebra as bilinears of the Gamma-matrices are in agreement
with a similar expression in \cite{Castellani}.
In addition, up to overall rescalings, these can be assembled in
\be
J^{ab}=\G^{ab}+{1\ov 4} \psi_{abcd} \G^{cd}\ ,\qq J^a=\G^{a+}\ ,
\label{je4}
\ee
where $\psi_{abcd}$ is the 4-index $G_2$-invariant tensor constructed as
the dual of the octonionic structure
constants \cite{occt}.\footnote{For the comparison of \eqn{g22g} with
\eqn{je4}, it is necessary to split the index $a=(i,\hat i,7)$
and write the $G_2$-tensors according to
the decomposition ${\bf 7\to 3+3+1}$ \cite{Bilal} 
(In particular, see eq. (A.4)).}
Let us also note that
the projections \eqn{pr9}, together with \eqn{uuni}, imply that the Killing
spinor is indeed invariant under the action of the holonomy group.


\section{Conclusions}

In this paper we have presented a systematic approach to Ricci-null manifolds with reduced
Lorentzian holonomy group. The manifolds that we have constructed are supersymmetric waves
in six, eight and nine dimensions, with a transversal Riemannian metric and
explicit dependence
on the light-cone time. An interesting feature of our metrics, when the temporal dependence arises
from variable moduli parameters, is that as the light-cone
time evolves their isometry group can be modified.

We have only covered purely gravitational solutions to the
equations of motion of
eleven-dimensional supergravity.
A more precise understanding of M-theory requires a complete picture
of all possible supersymmetric vacua, and must include configurations with non-vanishing
four-form flux. In the presence of background flux an important differences arises:
Killing spinors are no longer covariantly constant with respect to the Levi--Civita connection,
as in the purely gravitational case, but with respect to a connection on the spin bundle, which
gives rise to a generalized holonomy group \cite{Duff,Hullgh}. In fact, supersymmetric solutions
with non-vanishing four-form flux are not Ricci-flat, and have already been 
classified (see, for instance, \cite{Gauntlett1,Gauntlett2,Batrachenko}).
An interesting extension of the present work is the deformation
of our solutions to include a non-vanishing four-form flux, along the lines of \cite{Hull,FOF}.
The additional constraint arising from the equations of motion in this case simply implies that
the field strength four-form must be null.
It is of interest to explore further our method by constructing
more general solutions than the ones we
have presented here including, in particular, cases where fluxes are turned on.


\bs\bs

\centerline {\bf Acknowledgments}

R.H. wishes to thank M. Blau for discussions. In addition, he 
acknowledges the financial support provided through the European
Community's Human Potential Programme under contract HPRN-CT-2000-00131
``Quantum Structure of Space-time'', the Swiss Office for Education
and Science and the Swiss National Science Foundation.

K.S. acknowledges the financial support provided through the European
Community's Human Potential Programme under contracts HPRN-CT-2000-00131
``Quantum Structure of Space-time'' and
HPRN-CT-2000-00122 ``Superstring Theory'', by the Greek State
Scholarships Foundation under the contract IKYDA-2001/22 ``Quantum Fields
and Strings'', as well as NATO support by a Collaborative 
Linkage Grant under the
contract PST.CLG.978785 ``Algebraic and Geometrical Aspects of Conformal
Field Theories and Superstrings'' and support by the
INTAS contract 03-51-6346. 

D.Z. acknowledges
the financial support provided through the Research Committee of
the University of Patras for a ``K.Karatheodory'' fellowship under
contract number 3022.


\renewcommand{\theequation}{\thesection.\arabic{equation}}
\csname @addtoreset\endcsname{equation}{section}

\appendix

\section{A general example in six dimensions}

So far we have worked out explicit
examples in various dimensions where the minimal
choice \eqn{dkj11} turned out to be valid. However, this is not always possible
as we have seen in subsection 3.3.
Let us consider now a six-dimensional class of metrics based on
four-dimensional self-dual metrics with a translational Killing vector
field $\del/\del\tau$. The general form of the metric is \cite{Taub-NUT}
\be
ds_4^2= V(d\tau + \om_i dx^i)^2 + V^{-1} dx_i^2 \ ,\qq i=1,2,3\ ,
\label{multi}
\ee
where the $V$ and $\om_i$, $i=1,2,3$ obey
\be
\del_i V^{-1} =-\e_{ijk}\del_j \om_k\quad \Longrightarrow
\quad \del^2 V^{-1} =0 \ .
\label{jf0}
\ee
Hence the most general multi-center solution for $V^{-1}$ is
\be
V^{-1} =V_0+ \sum_{a=1}^N {m\ov |\vec x-\vec x_a|}\ .
\label{dke}
\ee
With the above choice, the singularities of the harmonic
function $V^{-1}$ at $\vec x={\vec x}_a$ correspond
to removable {\it nut} singularities of the metric \eqn{multi}
provided that the variable $\tau$
has period $4\pi m$. This is the reason that the strength $m$ for all of these
singularities has be chosen to be the same.
Hence, it follows that if the constant $V_0\neq 0$ (in which case it can
be normalized to 1), the space is asymptotically locally flat (ALF).
If $V_0=0$ then the space is asymptotically locally Euclidean (ALE), with
boundary at infinity $S^3/R_N$, where $R_N$ is a discrete subgroup of $SO(4)$.
The case of Taub--NUT we considered before corresponds to the one-center
ALF space and that of the Eguchi--Hanson to the two-center ALE space.
This solution preserves $1/2$ of the maximal supersymmetry provided
that the Killing spinor is constant and subject to the projection \eqn{pr4}.
Also, depending on the arrangement of the centers at $\vec x=\vec x_a$,
part of the symmetry of the $\mathbb{R}^3$ space spanned by the $x^i$'s
is preserved by the metric \eqn{multi}.

According to \eqn{D}, the ansatz for the six-dimensional
metric with $SU(2) \ltimes \mathbb{R}^4$ holonomy is
\be
d\hat{s}_6^2 =  2 du\, dv + 2 (V_i dx^i+V_{\tau} d\tau) du
+ F du^2 + V ( d\tau + \om_i dx^{i})^2 + V^{-1} dx_i^2 \ ,
\label{DDD}
\ee
where $F$, $V_{\tau}$ and $V_i$ are functions of $u$ and of the $x^i$'s.
It is easy to see that in trying to introduce a $u$-dependence,
the minimal choice \eqn{dkj11} is not enough, since
the condition \eqn{ee} is not satisfied. Hence, we have to try and employ
our method in
full generality. We will use the frame basis
\be
e^i= V^{-1/2} dx^i\ ,\quad i=1,2,3\ , \qq e^4 =V^{1/2} (d\tau + \om_i dx^i) \ ,
\ee
and for convenience parameterize $V_{\tau}$, $V_i$ and $F$ in
terms of the functions $\L$, $H$ and $\L_i$ whose properties are easy to
state
\be
F = \L + VH^2\ ,\qq V_{\tau}=VH\ ,\qq V_i = \L_i + VH\om_i \ ,\quad
i=1,2,3\ .
\ee
These auxiliary functions will be determined in order to satisfy the
constraints \eqn{jdk3}, \eqn{jh9} and \eqn{ddFg}.
We find that, $\L$ and $\L_i$, like $V^{-1}$,
are harmonic functions in $\mathbb{R}^3$, namely
\be
\del^2 \L = \del^2\L_i = 0 \
\label{jf4}
\ee
and in addition
\be
{dV^{-1}\ov du} = \del_i \L_i \ .
\label{jf2}
\ee
The function $H$ is then determined from
\be
\del_i H = \e_{ijk}\del_j \L_k + \dot \om_i \ ,
\label{hse}
\ee
whose integrability is guaranteed from \eqn{jf0}, \eqn{jf4} and \eqn{jf2}.
Then computing the matrix $\L^{ab}$ from \eqn{jdk3} we find
\be
\L^{i4}=H\del_iV + V \e_{ijk}\del_j \L_k \ ,\qq
\L^{ij}=V \del_{[i}\L_{j]} + H \e_{ijk}\del_k V\ .
\ee
The fact that $\L^{4i}+\ha \e_{ijk} \L^{jk}=0$, together with the projection
\eqn{pr4}, forces the condition \eqn{jh9} to be satisfied.

The harmonic functions $\L$ and $\L_i$
have forms similar to \eqn{dke}, with the
important observation that the $\L_i$'s are
linked to $V^{-1}$, as dictated by the
condition \eqn{jf2}. Taking this into account we summarize the result,
\be
V^{-1} =V_0+ \sum_{a=1}^N {m\ov |\vec x-\vec x_a(u)|}\ ,\qq
\L_i = -\sum_{a=1}^N {m \ \dot x^i_a(u) \ov |\vec x-\vec x_a(u)|}\ .
\ee
Hence, we see that the fixed centers at $\vec x=\vec x_a$ can move with
the (light-cone time) by becoming $u$-dependent. In addition
\be
\L = \sum_{a=1}^N {\l_a(u)\ov |\vec x-\vec y_a(u)|}\ ,
\ee
where $\vec x=\vec y_a$ represent a set of, in principle,
$u$-dependent centers, with the strength of each center denoted by
the arbitrary functions $\l_a(u)$.
We have also set the additive constants in $\L$ and $\L_i$ to zero
since they can be always absorbed by shifts of the coordinate $v$ in \eqn{D}.
Finally, the functions $\om_i$ and
$H$ are computed using \eqn{jf0} and \eqn{hse}, respectively.

In order to investigate the properties of our metric \eqn{DDD}, we have to
specify the kind of motion of the centers of the harmonic functions.
Let's consider the important class of such motions where the
centers of $V^{-1}$
assume definite fixed values at the remote past and the remote future
light-cone time $u$,
when also the strengths of the centers of $\L$ vanish. Namely,
\be
\lim_{u\to \pm \infty} x_a(u)=x^\pm_a\ ,\quad \lim_{u\to \pm \infty}
\dot x_a(u)=0\ , \quad \lim_{u\to \pm \infty} \l_a(u)=0\ ,
\qq a=1,2,\cdots , N\ .
\ee
Then, regularity of the solution near the centers in the remote past and
feature requires that they are distinct and all have strength $m$.
Therefore, our six-dimensional metric interpolates between two, generally
different, multi-center
metrics of the form \eqn{multi} (times the two-dimensional light-cone). Since
the arrangements of these centers could be different, the symmetries preserved
by the solution can also differ as well.



\begin{thebibliography}{99}

\renewcommand{\baselinestretch}{1}
\normalsize

\bibitem{Berger} M.~Berger, Bull. Soc. Math. France {\bf 83} (1955) 225.

\bibitem{Bryant} R.L.~Bryant,
in {\it Global analysis and harmonic analysis} (Marseille-Luminy, 1999),
53-94, S\'emin. Congr., 4, Soc. Math. France, Paris, 2000, {\tt math.dg/0004073}.

\bibitem{FOF} J.M.~Figueroa-O'Farrill,
Class.\ Quant.\ Grav.\  {\bf 17} (2000) 2925, {\tt hep-th/9904124}.

\bibitem{Hull} C.M.~Hull,
Phys. Lett. {\bf B139} (1984) 39.

\bibitem{Mueller}
M.~Mueller,
Nucl. Phys. {\bf B337} (1990) 37.

\bibitem{Greene}
B.R.~Greene, A.D.~Shapere, C.~Vafa and S.T.~Yau,
Nucl. Phys. {\bf B337} (1990) 1.

\bibitem{Kiritsis}
E.~Kiritsis and C.~Kounnas,
Phys. Lett. {\bf B331} (1994) 51,
{\tt hep-th/9404092}.

\bibitem{Tseytlin} A.A.~Tseytlin,
Phys. Lett. {\bf B334} (1994) 315
{\tt hep-th/9404191}.

\bibitem{GiPo} G.W.~Gibbons and C.N.~Pope,
Commun. Math. Phys.  {\bf 66} (1979) 267.

\bibitem{intersections1} A.~A.~Tseytlin,
Nucl.\ Phys.\ B {\bf 475} (1996) 149, {\tt hep-th/9604035}.

\bibitem{intersections2} E.~Bergshoeff, M.~de Roo, E.~Eyras, B.~Janssen and 
J.P.~van der Schaar,
Class.\ Quant.\ Grav.\  {\bf 14} (1997) 2757, {\tt hep-th/9704120}.

\bibitem{EH} T.~Eguchi and A.J.~Hanson,
Phys. Lett. {\bf B74} (1978) 249.

\bibitem{Taub-NUT} S.W.~Hawking, Phys. Lett. {\bf A60} (1977) 81.\hfill\break
G.W.~Gibbons and S.W.~Hawking, Phys. Lett. {\bf B78} (1978) 430.\hfill\break
G.W.~Gibbons and S.W.~Hawking,
Commun. Math. Phys. {\bf 66} (1979) 291.\hfill\break
T.~Eguchi, P.B.~Gilkey and A.J.~Hanson, Phys. Rept. {\bf 66} (1980) 213.

\bibitem{Batrachenko} A.~Batrachenko, M.J.~Duff, J.T.~Liu and W.Y.~Wen,
{\it Generalized holonomy of M-theory vacua},
{\tt hep-th/0312165}.

\bibitem{EN} J.D.~Edelstein and C. N\'u\~nez,
JHEP {\bf 0104} (2001) 028, {\tt hep-th/0103167}.

\bibitem{conifold} P.~Candelas and X.C.~de la Ossa,
Nucl. Phys. {\bf B342} (1990) 246.\hfill\break
L.A.~Pando Zayas and A.A.~Tseytlin,
JHEP {\bf 0011} (2000) 028, {\tt hep-th/0010088}.

\bibitem{Cvetic}
M.~Cvetic, G.W.~Gibbons, H.~Lu and C.N.~Pope,
Phys. Lett. {\bf B534} (2002) 172, {\tt hep-th/0112138}.

\bibitem{HS} R.~Hern\'andez and K.~Sfetsos,
Phys.\ Lett. {\bf B536} (2002) 294, {\tt hep-th/0202135}.

\bibitem{EPR1} J.D.~Edelstein, A.~Paredes and A.V.~Ramallo,
JHEP {\bf 0301} (2003) 011, \hfill\break
{\tt hep-th/0211203}.

\bibitem{BS} R. Bryant and S. Salamon,
Duke Math. J. {\bf 58} (1989) 829.\hfill\break
G.W.~Gibbons, D.N.~Page and C.N.~Pope,
Commun.\ Math.\ Phys.\  {\bf 127} (1990) 529.

\bibitem{BGGG} A.~Brandhuber, J.~Gomis, S.S.~Gubser and S.~Gukov,
Nucl. Phys. {\bf B611} (2001) 179, {\tt hep-th/0106034}.

\bibitem{Castellani} L.~Castellani and L.J.~Romans,
Nucl. Phys. {\bf B238} (1984) 683.

\bibitem{occt} M. G\"unaydin and F. G\"ursey, J. Math. Phys. {\bf 14} (1973) 1651.
\hfill\break
F. G\"ursey and C.-H. Tze, Phys. Lett. {\bf 127B}  (1983) 191.
\hfill\break
B. de Wit and H. Nicolai, Nucl. Phys. {\bf B231} (1984) 506.

\bibitem{Bilal} A.~Bilal, J.P.~Derendinger and K.~Sfetsos,
Nucl. Phys. {\bf B628} (2002) 112, \hfill\break
{\tt hep-th/0111274}.

\bibitem{Duff} M.J.~Duff and J.T.~Liu,
Nucl. Phys. {\bf B674} (2003) 217, {\tt hep-th/0303140}.

\bibitem{Hullgh} C.~Hull, {\em Holonomy and symmetry in M-theory},
{\tt hep-th/0305039}.

\bibitem{Gauntlett1} J.P.~Gauntlett and S.~Pakis,
JHEP {\bf 0304} (2003) 039, {\tt hep-th/0212008}.

\bibitem{Gauntlett2} J.P.~Gauntlett, J.B.~Gutowski and S.~Pakis,
JHEP {\bf 0312} (2003) 049, \hfill\break
{\tt hep-th/0311112}.

\end{thebibliography}
\end{document}